\documentstyle[prl,tighten,aps,epsf,graphics,multicol,times,bbm]{revtex}

\newcommand{\ket}[1]{\left | \, #1 \right \rangle}
\newcommand{\bra}[1]{\left \langle \, #1 \right |}
\newcommand{\be}{\begin{equation}}
\newcommand{\ee}{\end{equation}}
\newcommand{\bea}{\begin{eqnarray}}
\newcommand{\eea}{\end{eqnarray}}
\newtheorem{theorem}{Theorem}
\newtheorem{lemma}[theorem]{Lemma}

\newtheorem{conjecture}[theorem]{Conjecture}
\newcommand{\proofend}{\hfill\rule{2mm}{2mm}\medskip}

\begin{document}

\title{Optimal local implementation of non-local quantum gates}
\author{J. Eisert$^{1}$, K. Jacobs$^{2}$, P. Papadopoulos$^{3}$
and M.B. Plenio$^{3}$}
\address{1 Institut f{\"u}r Physik, Universit{\"a}t Potsdam, 14469 Potsdam, Germany}
\address{2 T--8, Theoretical Division,
Los Alamos National Laboratory, Los Alamos, New Mexico, 87545 USA}
\address{3 Optics Section, The Blackett Laboratory, Imperial College,
London SW7 2BW, England}

\maketitle

\begin{abstract}
We investigate the minimal resources that are required in the
local implementation of non-local quantum gates in a distributed
quantum computer. Both classical communication requirements and
entanglement consumption are investigated. We present general
statements  on the minimal resource requirements and present
optimal procedures for a number of important gates, including CNOT
and Toffoli gates. We show that one bit of classical communication
in each direction is both necessary and sufficient for the
non-local implementation of the quantum CNOT, while in general two
bits in each direction is required for the implementation of a
general two bit quantum gate. In particular, the state-swapper
requires this maximum classical communication overhead. Extensions
of these ideas to multi-party gates are presented.\\

PACS: 03.67.-a
\end{abstract}

\begin{multicols}{2}

\section{Introduction}
A quantum computer \cite{Barenco96,Plenio98,Steane98} allows, in
principle, for the efficient solution of some problems that are
intractable on a classical computer, the most striking example
being the factorization of large numbers \cite{Shor94,Jozsa96}.
However, the practical problems involved in the actual
construction of a quantum computer of an interesting size
(certainly more than 50 qubits) that is capable of performing a
sufficiently large number of logical gates (a few hundred appear
as a lower limit for an interesting problem involving 50 qubits)
are daunting. Problems range from fundamental effects such as
decoherence and dissipation, experimental imperfections for
example in the timing, length and intensity of the laser pulses to
the non-trivial task of storing and isolating reliably a large
number of qubits \cite{Steane98,Wineland,Chaos,Decoherence}. In
fact, in proposals such as ion trap or the cavity QED
implementations it seems problematic to store and process very
large numbers of qubits in a single `processor'. A possible way
out would be the construction of a quantum computer not as a local
device that contains all qubits in a single processor, but to
build it from the outset as a multi-processor device where each
processor contains only a small number of qubits. Such a
'distributed quantum computer' can be viewed as a generalization
of a quantum communication network in which each node can act as a
sender or receiver and contains only a small number of qubits.
Distributed quantum computation has been considered previously by
Grover \cite{Grover}, and he demonstrated that the solution of a
phase estimation problem can be obtained efficiently with such a
device assuming ideal conditions. It was later shown, that even
under non-ideal conditions, i.e., in the presence of decoherence, a
distributed quantum computer can be superior to a classical computer
in terms of the resources that are required for the solution of the
phase estimation problem \cite{Huelga}. However, these
investigations considered the specific problem of phase estimation
and did not address the question of universal quantum computation.
Before one is able to consider the physical resource efficiency of
a distributed quantum computer in general, it is necessary to
establish first optimal implementations of quantum gates between
qubits that are located in different nodes of the distributed
quantum computer. This problem is addressed in this paper. We
present optimal protocols implementing gates that affect qubits in
different nodes (here dubbed non-local gates) only using local
operations and classical communication (LOCC) and previously
shared entanglement. Optimality is measured in terms of the
consumption of the basic experimental resources of entanglement
and classical communication between nodes. We present general
theorems that give lower bounds on the resources required for the
implementation of quantum gates and for several universal quantum
gates we present optimal implementations. We also discuss the
general structure of the classical communication transfer in these
implementations.

It should be noted that the issue addressed in the present paper is different
from the question as to whether (and how) a particular entanglement
transformation is possible under local quantum operations
and classical communication \cite{Nielsen} in that
in the course of the
non-local implementation of a quantum gate the
initial state is not known in advance. Instead,
with the use of shared entanglement particular joint
unitary operations between several parties are simulated.

In Section II we begin with an investigation of two-qubit gates.
We establish some lower bounds on the resources that are required
to implement two-qubit gates and present optimal implementations
for a number of important gates. In particular we present a
protocol that implements a CNOT gate consuming one ebit of
entanglement and using only one classical bit of communication
between the two parties. We then proceed in Section III to study
multi-party gates such as Toffoli gates and other more general
multi-party quantum gates again presenting bounds on the required
physical resources and optimal protocols for some important classes
of gates.

\section{Non-local two-qubit gates}
General single-bit rotations together with a CNOT gate are
sufficient to implement any multi-qubit unitary transformation.
This implies that the resource requirements for the implementation
of a CNOT gate are a limiting factor in the construction of general
unitary transformations in a distributed quantum computer. For this
reason we investigate first the CNOT gate.
\begin{theorem} \label{theorem1} One bit of classical communication in each direction and
one shared ebit is necessary and sufficient for the non-local
implementation of a quantum CNOT gate.
\end{theorem}
{\it Proof:} (i) {\em Necessity}: To demonstrate that one bit of
communication in each direction is necessary we first note that
the procedure consists of local operations and classical
communication. As local operations cannot transmit information
from Alice to Bob, or vice versa, all information which has been
sent at the end of the operation must have been sent classically.
Consider now the CNOT quantum gate. If the target qubit is
initialised in the state $|0\rangle$, then its final state will be
$|0\rangle$ or $|1\rangle$ depending on the initial state of the
control qubit being $|0\rangle$ or $|1\rangle$ respectively.
Therefore, the final result of the gate in this case is the
communication of one bit of information from Alice (holding the
control qubit) to Bob (holding the target qubit). Consequently, in
the non-local implementation, one bit of classical information
must have been sent classically from Alice to Bob. The reason for
this can be seen from an elegant argument presented in the figure
caption of the last figure in \cite{qtele} (see \cite{Polykarpos}
for more details). In short, assume that Alice needs to send less
than one bit. In that case she could omit sending the bit and
force Bob to make a guess. As he would guess the correct answer
with a probability larger than $1/2$, Alice and Bob could then use
error correction codes to establish a perfect channel and would
end up with a superluminal communication channel. To see that one
bit must also have been sent from Bob to Alice, we need merely
note that in the basis $|\pm\rangle = (|0\rangle \pm
|1\rangle)/\sqrt{2}$ the role of control and target in a CNOT gate
are reversed. Consequently, if Alice's  particle is prepared in
the standard state $|+\rangle$ and Bob chooses  to prepare his
particle either in state $|+\rangle$ or $|-\rangle$, Alice will,
after the application of the CNOT gate, hold a particle which is
either in state $|+\rangle$ or $|-\rangle$ depending on the state
Bob's  particle has been prepared in. Therefore one bit of
information has been transmitted from Bob to Alice. As the
implementation of the CNOT must be independent of the initial
state, the procedure must allow for one bit of communication in
each direction, and as a consequence the non-local implementation
must involve, as a minimum, one bit of communication in both
directions.

That one ebit is required can be seen from the fact
that a CNOT gate acting on the initial state
$(|0\rangle_A+|1\rangle_A)|0\rangle_B$ leads to a maximally entangled
state $(|00\rangle_{AB}+|11\rangle_{AB}$. As the amount of entanglement
cannot be increased by local operations, this implies that the
non-local implementation of a CNOT gate must consume at least one
ebit.

\begin{figure}
\leavevmode \epsfxsize=8cm \epsfbox{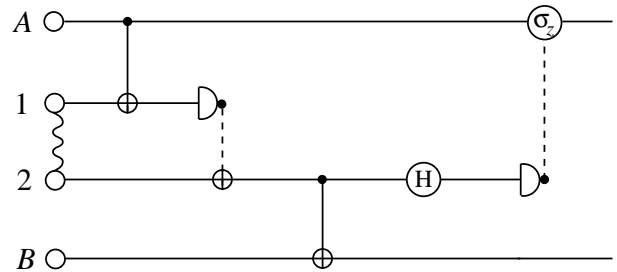} \vspace{0.3cm}
\caption{\narrowtext A quantum circuit to perform the CNOT
non-locally with minimal classical communication. Alice has the
qubits $A$ and $A_1$, and Bob has $B$ and $B_1$. Alice and Bob are
only allowed to communicate classically, and this communication is
represented by the dashed lines. Each dashed line denotes one bit
of communication.} \label{fig1}
\end{figure}

(ii) {\em Sufficiency}: In the following we construct a quantum
circuit which performs the CNOT non-locally using one e-bit and
the transmission of one classical bit in each direction. This
quantum circuit is given in figure~\ref{fig1}. The CNOT is
performed between the qubits $A$ and $B$. Alice holds the qubits $A$
and $A_1$, and Bob holds the qubits $B$ and $B_1$. The wavy line connecting
$A_1$ and $B_1$ signifies that they are entangled. In
particular we will choose their initial state to be
$(|00\rangle + |11\rangle)/\sqrt{2}$.
The initial state of $A$ is
necessarily arbitrary, and so is given by $\alpha |0\rangle_A +
\beta |1\rangle_A$. The initial state of $B$ is also arbitrary, and
is given by $\gamma |0\rangle_B + \delta |1\rangle_B$. Time now flows
from left to right in figure~\ref{fig1}. First a local CNOT is
performed with $A$ as the control and $A_1$ as the target. After this
the combined state of $A$, $A_1$ and $B_1$ is
\begin{equation}
    \frac{1}{\sqrt{2}}(\alpha |000\rangle + \alpha |011\rangle +
    \beta |110\rangle + \beta |101\rangle)_{A A_1 B_1}.
\end{equation}
Alice then performs a measurement on $A_1$ in the computational
basis, and  the line corresponding to this qubit terminates. The
result of the measurement is one bit of information, which is
communicated to Bob, and this communication is denoted by the
dashed line. If the result is $|0\rangle$ Bob does nothing, and if
the result is $|1\rangle$ Bob performs the not operation. At this
point the combined state of $A$ and $B_1$ is $\alpha
|00\rangle_{AB_1} + \beta |11\rangle_{AB_1}$. That is, we have now
effectively performed a CNOT between $A$ and $B_1$, in which the
initial state of $2$ was $|0\rangle$. Now particle $B_1$ contains
the necessary information about the state of $A$. We can now
perform a CNOT between $B_1$ and $B$. The combined state of $A$,
$B_1$ and $B$ is now
\begin{equation}
  \frac{1}{\sqrt{2}}(\alpha\gamma |000\rangle + \alpha\delta |001\rangle +
  \beta \delta|110\rangle + \beta\gamma |111\rangle)_{A B_1 B}.
\end{equation}
All we have to do  is to remove $B_1$ from the state. This is done
by performing a Hadamard transformation on $B_1$, and then
measuring $B_1$ in the computational basis, at which point the
line denoting $B_1$ terminates. The result of the measurement (one
bit) is communicated to Alice. If the result is '0' Alice does
nothing, and if the result is '1' she performs a
(state-independent) $\sigma_z$ operation on particle $A$. This
completes the non-local CNOT. \proofend

\begin{theorem} \label{theorem2}
A control-U gate can be implemented using one shared
ebit and one bit of classical communication in each direction.
\end{theorem}
{\it Proof:} A control-U gate is defined as a gate that applies
the identity on the target qubit if the control bit is in state
$|0\rangle$ and it applies the unitary operator $U$ to the target
if the control qubit is in state $|1\rangle$. The same quantum
circuit as in Fig. 1 can be used except that the CNOT gate on Bobs
side is replaced by a control-U gate. \proofend

In general a single application of a control-U gate cannot be
employed to create one e-bit from an initial product state of two
qubits. Furthermore, the amount of classical information that can
be sent from Alice to Bob via a general control-U gate is less
than one bit. This raises the question as to whether such a
control-U gate can be implemented with less resources than a full
ebit and one classical bit of communication in each direction.
Clearly this will not be possible when we only wish to implement a
single instance of a control-U gate. However, it may be
conceivable  that one has a situation in which one needs to carry
out a large number of control-U gates simultaneously. In that case
it is conceivable that this could be done with less than 1 ebit of
entanglement per gate and less than one bit of classical
communication in each direction. However, this turns out to be a
difficult question and we have been unable to find such a
scheme.\\

Let us now move on to investigate general two-qubit quantum
gates to establish the minimum resource requirements for their
implementation.
\begin{theorem} \label{theorem3}
Two bits of classical communication in both directions
and two shared ebits is sufficient for the non-local implementation
of a general two-bit gate.
\end{theorem}
{\it Proof:} To demonstrate that this amount
of communication is sufficient to
implement all quantum operations we need merely invoke quantum
teleportation. Any operation may be performed by teleporting
Alice's state to Bob, at which point Bob may locally perform the
operation, and then teleport the resulting state back to Alice.
This procedure requires two bits of communication in each
direction and $2$ shared ebits \cite{qtele,plenio} \proofend \\

\begin{figure}
\leavevmode
\centerline{\epsfxsize=5.0cm
\epsfbox{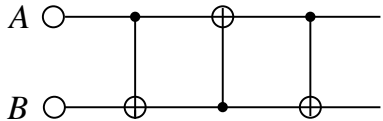}
}
\vspace{0.3cm}
\caption{\narrowtext A state swapper implemented by means
of three quantum CNOT gates. } \label{fig2}
\end{figure}

Moreover, there are two-qubit gates  that require two bits of
classical communication in each direction and consumes $2$ bits.
An example is the state-swapper, which may be written as three
CNOT gates, one after the other, with Alice as the control,
target, and then control, in that order (see Fig. \ref{fig2}) .
To show that two bits of classical communication are required
(each way) in the non-local implementation of this gate, we need
to show that this amount of information may be communicated from
Alice to Bob (and vice versa) when the gate is performed. To do
this we merely have to note that at the completion of the gate
Alice has sent her state to Bob. Now, this state could have been
initially in a maximally entangled state with a qubit that Bob
possesses. Dense coding tells us that this enables Alice to
send two bits of information to Bob~\cite{sdcoding}. Naturally, 
Bob can use 
the same procedure to send two bits of information to
 Alice. Therefore, in a non-local
implementation, the state swapper requires at least two bits of
communication in each direction. An analogous argument shows that
the state swapper would also require two shared ebits, as a state
swapper can be used to establish two ebits from a product state.
To achieve this one simply applies the state swapper to particles
$A_2$ and $B_2$ of the state
$(|00\rangle_{A_1A_2}+|11\rangle_{A_1A_2})(|00\rangle_{B_1B_2}+|11\rangle_{B_1B_2})$.

It is remarkable that the swap gate requires only two shared ebits
as it can be shown that three CNOT gates are necessary to
implement it when one employs the ordinary gate array picture
using a universal set of quantum gates that is made up of CNOT
gates and local unitary operations \cite{Pleniounpub}. This
observation may be useful, as it demonstrates that in some cases the
use of entanglement can be replaced partially by local
measurements and classical communication.

Before we move on to investigate the implementation of non-local
multi-party gates we would like to analyze the structure of the classical information
transfer involved in the gate implementation somewhat
further. In both examples discussed above it turned
out that the classical information
transfer between the two parties is symmetric, i.e., the same
number of bits need to be sent from Alice to Bob and vice versa.
Likewise, the amount of classical information that can be sent
using these two-qubit gates is also the
same in each direction. It is therefore quite natural to ask
whether this is the case in general.
Indeed we have not been able to find a counter-example and we
therefore make the following two closely related conjectures.

\begin{conjecture} \label{Proposition1} The minimal amount of classical
communication required to implement any two-party quantum gate
with one qubit associated with each party and shared $M$ ebits,
$M=1,2$, is always the same in each direction.
\end{conjecture}

\begin{conjecture} \label{Proposition1a}
The amount of classical information that can be
sent via any two-qubit gate is the same in each direction.
\end{conjecture}

\begin{figure}
\leavevmode \epsfxsize=8.2cm \epsfbox{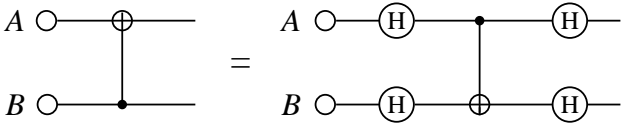}
\vspace{0.3cm} \caption{\narrowtext A CNOT gate, with $A$ as control
and $B$ as target, surrounded by Hadamard gates is equivalent to a
CNOT gate with $A$ as target and $B$ as control.} \label{CNOTswap}
\end{figure}

While these conjectures appear natural, we have not been able to
find general proofs for them. However, we have been able to
confirm both of them for a number of classes of two-qubit quantum
gates. An example of a gate which has the same classical
information capacity in both directions is the CNOT gate whose
optimal implementation has been described above. How can we see
that a quantum gate is symmetric with respect to its capability
for classical information transfer? Before we move on to the most
general case, let us consider the CNOT gate. Imagine we have the
ability to perform a CNOT gate with Alice as the control and Bob
as the target. Using this gate and local operations only, we can
then also implement a CNOT with Alice as a target and Bob as a
control, simply by applying a Hadamard gate to each qubit both
before and after the CNOT, see Fig. \ref{CNOTswap}.
The two versions of the CNOT gate are also related via the
(nonlocal) state swapper.
\begin{eqnarray}\label{CNOTswapping}
    U_{\text{CNOT}}^{BA} &=& 
	 U_{ss} U_{\text{CNOT}}^{AB} U_{ss}^{\dagger}\nonumber \\
	 & = &(H\otimes H)U_{\text{CNOT}}^{AB} (H\otimes H).
\end{eqnarray}
where $U_{\text{CNOT}}^{AB}$ represents the CNOT gate with $A$ as a
control and $B$ as a target and $U_{ss}$ denotes the state
swapper.  In general  if we can achieve the transformation
 $U_{BA}\equiv U_{ss} U_{AB} U_{ss}^{\dagger}$
 from $U_{AB}$ and purely local operations, i.e., if
there exist local one-qubit unitary operators $U_1$, $U_2$, $U_3$
and $U_4$ for which we have
\begin{eqnarray}\label{condition}
 U_{BA} &= &U_{ss} U_{AB} U_{ss}^{\dagger}\nonumber\\
  &=& (U_1\otimes U_2) U_{AB} (U_3\otimes
 U_4)
 \end{eqnarray}
then Eq.\ (\ref{condition}) is a
sufficient condition for the classical information transmission
capacities in each direction to be equal. In the following we will
determine some sets of quantum gates $U_{AB}$ for which Eq.\
(\ref{condition}) holds.

Let us begin with a slightly simpler problem.
Suppose that we have a two-qubit quantum gate $V_1\in U(4)$. $V_1$
can be expressed in terms of its generator as $V_1=\exp({iH_1})$,
where the generator $H_1$ is a Hermitean operator. We now define
another quantum gate $V_2$ as
\begin{eqnarray}
 V_2\equiv
 U_{ss}V_1U_{ss}^{\dagger} &=& U_{ss}e^{iH_1}U_{ss}^{\dagger}\nonumber\\
 &=&e^{iU_{ss}H_1U_{ss}^{\dagger}}\equiv
 e^{iH_2}
\end{eqnarray}
where the generator $H_2$ of $V_2$ is clearly a Hermitean
operator. Our goal can therefore be reformulated as:
For which unitary operators $V_1$ can we write $V_2$ as $V_2=
(U_1\otimes U_2)V_1(U_1^{\dag}\otimes U_2^{\dag})$, or
equivalently for which generators $H_1$ of $V_1$ can we write
\begin{equation}\label{H}
 H_2\equiv U_{ss}H_1 U_{ss}^{\dagger}= (U_1\otimes U_2)H_1
 (U_1^{\dag}\otimes U_2^{\dag}).
\end{equation}
Note that this is less general than the transformation in Eq.
(\ref{condition}). It is useful to realize that both the unitary
operator $V_1$ and its generator
$H_1$ are diagonal in the same basis, say $\{\ket{\phi_i}, i=1,\;
2,\; 3,\; 4\}$. Furthermore, we can decompose $H_1$ with respect
to its eigenvectors as $H_1=\sum_i \lambda_i
\ket{\phi_i}\bra{\phi_i}\equiv\sum_i \lambda_i \rho_i$, where
$\lambda_i$ is the eigenvalue of $H_1$ corresponding to the
eigenvector $\ket{\phi_i}$. Consequently, Eq.\ (\ref{H})
becomes
\begin{equation}\label{a}
 \sum_i \lambda_i U_{ss}\rho_i U_{ss}^{\dagger}=
 \sum_i \lambda_i (U_1\otimes U_2) \rho_i
 (U_1^{\dag}\otimes U_2^{\dag})
\end{equation}
We can now prove a number of lemmas. We begin with
\begin{lemma} \label{lemma1}
Any two-qubit quantum gate that has a generator with
a single non-vanishing eigenvalue is symmetric with respect to
its classical information transfer capacity.
\end{lemma}
{\it Proof:} Suppose that the \textit{only} non-vanishing eigenvalue
of the generator $H_1$ is $\lambda_1$ \cite{CNOT}. In that case we
can always find one-qubit unitary operators $U_1$ and $U_2$ such
that Eq.\ (\ref{a}) holds. To see this, note that the eigenstate
$\ket{\phi_i}$ is actually a pure state describing a system composed
by two qubits. Therefore, it has the Schmidt decomposition
$\ket{\phi_1}=\sum_k \sqrt{p_k} \ket{k}_A \tilde{\ket{k}}_B\equiv \sum_k \sqrt{p_k} \ket{k}
\tilde{\ket{k}}$. Furthermore, in this case we have
\bea
  \sum_i&\lambda_i& U_{ss}\ket{\phi_i}\bra{\phi_i}
  U_{ss}^{\dagger}=\lambda_1 \sum_{k, l} \sqrt{p_k p_l} \tilde{\ket{k}} \ket{k}
  \tilde{\bra{l}}
  \bra{l}\nonumber\\=({\tilde{U}}&\otimes& U) \left(\sum_{k, l} \lambda_1 \sqrt{p_k p_l}
  \ket{k}  \tilde{\ket{k}}   \bra{l}
  \tilde{\bra{l}}\right)
  ({\tilde{U}}^{\dag}\otimes U^{\dag})\nonumber\\
  =({\tilde{U}}&\otimes& U) \left(\sum_i\lambda_i
  \ket{\phi_i}\bra{\phi_i}\right)
  ({\tilde{U}}^{\dag}\otimes U^{\dag}),
\eea
where $U$ is defined to be the unitary operator which maps each
basis vector $\ket{i}$ to its corresponding $\tilde{\ket{i}}$.
Similarly, the unitary operator $\tilde{U}$ maps each basis vector
$\tilde{\ket{i}}$ to its corresponding $\ket{i}$, i.e.,
$\tilde{U}=U^{\dag}$. 
\proofend

Another non-trivial class of quantum gates $U_{bd}$ for which condition
(\ref{H}) holds, is the one whose generator is Bell diagonal, i.e., we have
\begin{lemma} \label{lemma4}
Any two-qubit quantum gate that has a generator which
is Bell-diagonal is symmetric with respect to
its classical information transfer capacity.
\end{lemma}
{\it Proof:} If $\ket{\Psi}$ is \textit{any} of the Bell states,
the reader can easily verify that
$$
 \ket{\Psi}\bra{\Psi}=U_{ss} \ket{\Psi}\bra{\Psi} U_{ss}^{\dagger}=
 (\sigma_z \otimes \sigma_z)
 \ket{\Psi}\bra{\Psi} (\sigma_z \otimes \sigma_z)
$$
Therefore, for the quantum gate $U_{bd}$, condition (\ref{H}) is
satisfied by either choosing $U_1=U_2={\mathbbm{1}}$ or $U_1=U_2=\sigma_z$.
Recall that $\sigma_z$ is the Pauli matrix corresponding to the
arbitrarily chosen z direction. \proofend

Note however, that condition (\ref{H}) is not satisfied for \textit{all}
quantum gates $U_{AB}$. A counterexample is the gate
\begin{eqnarray}
 U_{AB}&=&e^{i\lambda_1} \ket{0+}\bra{0+} +
 e^{i\lambda_2} \ket{0-}\bra{0-}\nonumber\\
 &+& e^{i\lambda_3} \ket{10}\bra{10}+
 e^{i\lambda_4} \ket{11}\bra{11}.
\end{eqnarray}
For $\lambda_1=\lambda_2=0$ and non-trivial choice of $\lambda_3$
and $\lambda_4$ it is \textit{not} possible to find local unitary
operators $U_1$ and $U_2$ such that Eq.\ (\ref{H}) is
satisfied. Nevertheless, it is possible to find local unitary
operators $U_1$, $U_2$, $U_3$ and $U_4$ which satisfy the more
general condition (\ref{condition}). The local unitary operators
will be of the form \cite{John}:
\begin{eqnarray}
 &&U_1=e^{-i\lambda_4}\ket{1}\bra{1}+e^{i(\lambda_3-\lambda_4)}\ket{0}\bra{0},\\
 &&U_2=\ket{1}\bra{1}+e^{-i(\lambda_3-\lambda_4)}\ket{0}\bra{0},\\
 &&U_3={\mathbbm{1}},\\
 &&U_4=e^{i\lambda_4}\ket{1}\bra{1}+\ket{0}\bra{0}.
\end{eqnarray}
We can then conclude to the following lemma:
\begin{lemma} \label{lemma5}
 The amount of classical information that can be sent via any control-U
 gate of the form $$U=\ket{0}\bra{0}\otimes  {\mathbbm{1}} +
 \ket{1}\bra{1}\otimes\left(e^{i\lambda_3}\ket{0}\bra{0} +
 e^{i\lambda_4}\ket{1}\bra{1}\right)$$ is the same in each direction.
\end{lemma}
It should be noted that this does not mean that the amount of
information transferred in any particular operation of the gate
will be the same in both directions, as this will depend upon the
choice of initial states. However, an implementation of the gate
must work for all possible initial states, (in particular it must
work for the case where both qubits are pure and therefore contain
their maximum capacity), and this is what puts the limit on the
minimal communication requirement.

It is clear that we may now put 2-bit quantum gates into two
classes. Those which require no more than one bit of two-way
communication, and those that require more than one bit (but no
more than two bits). The CNOT falls into the first category, and
the state-swapper falls into the second. Two other standard gates
which fall into the first category are the c-U (which performs a
unitary transformation on one system depending on the state of the
other), and the state-preparer.

\section{Non-local multi-party gates}
In the previous section we have presented a number of results
concerning the implementation of non-local two-qubit quantum
gates in a distributed quantum computer. In the following we
will generalize these ideas to local implementation of multi-qubit
gates, i.e., gates where more than two parties are involved.
To illuminate the system behind the construction, we explain the
implementation of the Toffoli gate from which the generalization
to other multi-party gates will be evident.

\begin{theorem} \label{theorem4}
Two shared ebits and a total of four bits of classical communication
are necessary and sufficient for the local
implementation of a non-local three-party quantum Toffoli gate.
\end{theorem}

{\it Proof:} (i) Necessity:   A Toffoli gate can be reduced to an
ordinary CNOT gate when one fixes the state of one of the control
qubits to be $|1\rangle$. Chose the state of party $A$ to be
$|1\rangle$. Then the initial state is
\begin{equation}
    |\psi_{\text{ini}}\rangle = |1\rangle_A (\alpha |0\rangle + \beta
    |1\rangle) (\gamma |0\rangle + \delta |1\rangle),
\end{equation}
and after the application of the Toffoli gate we find
\begin{equation}
    |\psi_{\text{ini}}\rangle = |1\rangle_A (\alpha\gamma |00\rangle + \alpha\delta|01\rangle
    + \beta\gamma |11\rangle + \beta\delta |10\rangle)_{BC},
\end{equation}
which shows that we have implemented a CNOT between parties $B$ and
$C$. Therefore, Theorem \ref{theorem1} implies that one classical bit has to be
exchanged in both directions between $A$ and the target party $C$ and one
ebit has to be shared between them.
The same argument applies when we fix the state of qubit $B$
to be $|1\rangle$.

\begin{figure}
\leavevmode \epsfxsize=8.2cm \epsfbox{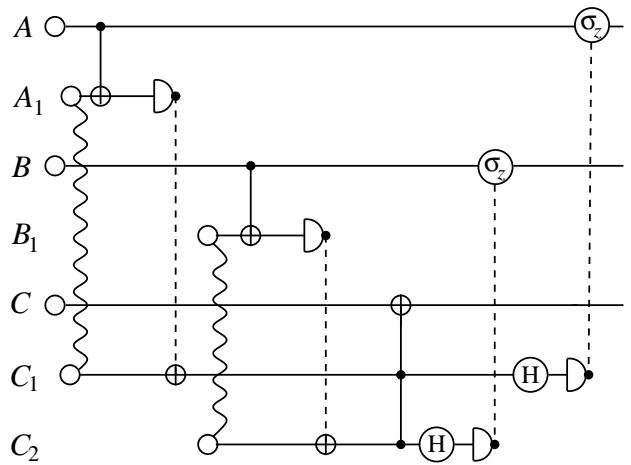} \vspace{0.3cm}
\caption{\narrowtext A quantum circuit for the non-local
implementation of a Toffoli gate. } \label{figtof}
\end{figure}

{\em Sufficiency.} The implementation of the Toffoli gate with
these minimal resources is presented in Fig. \ref{figtof}. Assume
that Alice and Clare share a pair $A_1$, $C_1$ of qubits in a
maximally entangled state $|\phi^+\rangle =
(|00\rangle+|11\rangle)/\sqrt{2}$, and that Bob and Clare share
another pair of particles $B_1$ and $C_2$ in the same state. Then
the initial state of the whole system consisting of particles $A$,
$B$, $C$, $A_1$, $B_1$, $C_1$ and $C_2$ is of the form
\begin{equation}
    |\psi\rangle=
    |\psi\rangle_A\otimes |\psi\rangle_B\otimes  |\psi\rangle_C
    \otimes |\phi^+\rangle_{A_1 C_1}\otimes |\phi^+\rangle_{B_1 C_2},
\end{equation}
where
\begin{eqnarray}
    |\psi\rangle_A&=& \alpha|0\rangle+\beta|1\rangle,\\
    |\psi\rangle_B&=& \gamma|0\rangle+\delta|1\rangle,\\
    |\psi\rangle_C&=& \eta|0\rangle+\xi|1\rangle.
\end{eqnarray}
The first step is a local quantum CNOT gate on $A$ and $A_1$ with
$A$ as control. Then Alice measures particle $A_1$ and Clare
performs a NOT operation on her particle $C_1$ if Alice finds
$|1\rangle$ and the identity if Alice finds $|0\rangle$. Qubit
$A_1$ is subsequently discarded. Now Bob applies a local CNOT with
$B$ being the control and $B_1$ being the target. Then Bob
measures particle $B_1$ and Clare performs a NOT operation on her
particle $C_2$ if Bob finds $|1\rangle$ and the identity if Bob
finds $|0\rangle$. Qubit $B_1$ is subsequently discarded. Now the
state of the remaining qubits $A$, $B$, $C$, $C_1$ and $C_2$ is
given by
\begin{eqnarray}\label{stage}
    (\alpha|00\rangle+\beta|11\rangle)_{AC_1}\otimes
    (\gamma|00\rangle + \delta |11\rangle)_{BC_2} \otimes
    |\psi\rangle_C.
\end{eqnarray}

In a further step Clare applies locally a Toffoli with $C_1$ and
$C_2$ being the control qubits. Subsequently  Clare applies
Hadamard gates to the qubits $C_1$ and $C_2$. Then she measures
$C_2$ and applies $\sigma_z$ or the identity $\bf 1$ to $B$ if her
result is $|1\rangle$ or $|0\rangle $ respectively. Finally she
measures $C_1$ and applies $\sigma_z$ or the identity to $A$ if
her result is $|1\rangle$ or $|0\rangle $ respectively. This
completes the Toffoli gate. \proofend

The total number of classical bits which have to be communicated
is four, and only two shared ebits of entanglement
are consumed.
Again, these results can be generalized to three-party
control-U operations that can be represented
in matrix form with respect to the computational basis as
\begin{equation}\label{cu}
	{\mathbbm{1}}_6 \oplus 
    \left(
    \begin{array}{cc}
     u_{00} & u_{01}\\
     u_{10} & u_{11}\\
    \end{array}
    \right),
\end{equation}
where
\begin{equation}
    \left(
    \begin{array}{cc}
    u_{00} & u_{01}\\
    u_{10} & u_{11}
    \end{array}
    \right)
\end{equation}
is the matrix representation of a unitary operator $U$. We only need
to replace the local Toffoli gate by a local three-party
control-U. This gives rise to
\begin{lemma} \label{lemma6} A three party control-U gate can be
implemented using four bits of classical communication
and two shared ebits.
\end{lemma}
Using Theorem \ref{theorem4} and Lemma \ref{lemma6} we are now in
a position to construct every possible quantum gate array using
only ebits, classical communication and local operations. In
particular one could use the results in \cite{gates} to construct
$N$-party controlled gates from CNOTs and single bit rotations.
This, however, is not optimal in terms of physical resources.
While it will be difficult to construct the optimal procedure for
general quantum gates, for some gates we are able to find these
procedures. We find for example

\begin{theorem} \label{lemma7} An $N$ party control-U gate can be
implemented using $2(N-1)$ bits of classical communication
and $N-1$ shared ebits.
\end{theorem}

{\it Proof:}
The control parties are enumerated from $P_1$ to $P_{N-1}$ and
each of them is carrying one ancilla numerated by
$P'_1$ to $P'_{N-1}$. The target qubit is denoted by $T$ and the
target party  possesses $N-1$ further ancillary qubits.
The first $N-1$ steps of the protocol are essentially analogous.
In the k-th  step a local quantum CNOT gate is applied on $P_k$
and $P'_k$ with $P_k$ as control. Then this party measures
particle $P'_k$ and the target party performs a NOT operation on
her ancillary qubit $T_k$ if Alice finds $|1\rangle$ and the
identity if Alice finds $|0\rangle$. Qubit $P'_k$ is subsequently
discarded. Now   we apply an  $N$-party controlled U gate on
Clares particles, with the ancillas $C_1, \ldots, C_{N-1}$ being
the control qubits and $T$ the target. Subsequently the target
party performs Hadamard gates on each of its ancillas.

This is then followed by $N-1$ steps involving measurements. In
the k-th  step qubit $T_k$ is measured in the
$|0\rangle,|1\rangle$ basis. If the outcome is $|1\rangle$,  then
$\sigma_z$ is applied to the qubit $P_k$;  if the outcome is
$|0\rangle$ then no action is taken on qubit $P_k$. Qubit $T_k$ is
subsequently discarded. Hence, the total required resources are
$
    2(N-1) \text{ bits of classical information}$ and $N-1$
initially shared ebits. \proofend

The amount of consumed resources in the latter protocol is rather
surprizing. In an inefficient non-local implementation of the
above $N$-party gate one could employ the simulation of the gate
with the use of two-party control-U gates and CNOT gates as in
Ref.\ \cite{gates}, but such that each step is realized
non-locally. In such a procedure a supply of  $3\times 2^{N-1}-4$
ebits would be necessary.

\begin{figure}
\leavevmode \epsfxsize=8.2cm \epsfbox{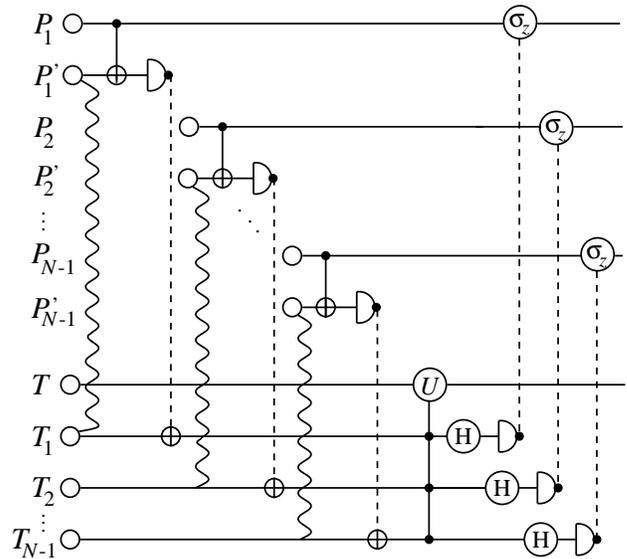} \vspace{0.3cm}
\caption{\narrowtext A quantum circuit for the non-local
implementation of an $N$-party control-U gate. } \label{fig3}
\end{figure}

A more efficient teleportation-based protocol \cite{Chefles} in
which the respective states of the qubits at different nodes are
twice teleported would still use $2(N-1)$ ebits and $4(N-1)$ bits
of classical information.

\section{Conclusions}
In this work we have addressed the problem of the local implementation
of non-local gates in a distributed quantum computer, i.e. a computer
which is composed of many subunits (local processors). Such a configuration
may be useful, as it requires only a small number of qubits (e.g. ions)
to be stored at each site which may be experimentally more feasible
than storing a large number of qubits in a single site. However, this
raised the issue of the non-local implementation of quantum gates.
We have addressed this question and have shown what the minimal resources
for the implementation of two-qubit quantum gates are. We have presented
explicit optimal constructions for the local implementation of non-local
control-U gates. We have generalized these results to multi-party
gates such as for example the Toffoli gate. We have also adressed some
issues concerning the structure of the information exchange that is required in these
implementations. We hope that this work will be useful for the assessment
of the viability of distributed quantum computation. \\

We acknowledge useful discussion with Daniel Jonathan and John Vaccaro.
This work was supported by the Deutsche Forschungsgemeinschaft (DFG),
the UK engineering and physical sciences research council (EPSRC),
The Leverhulme Trust, the European Science Foundation (ESF) programme
on quantum information processing, the EQUIP programme of the European
Union and the State Scholarships Foundation of Greece.

{\em Endnote:} During completion of this work we became aware of
the closely related work by D. Collins, N. Linden, and S. Popescu,
Phys. Rev. A {\bf 64}, 032302 (2001),
quant-ph/0005102.

\end{multicols}


\begin{references}
\bibitem{CNOT} A characteristic example for this case is the CNOT.
If $\{\ket{00}, \ket{01}, \ket{10},
\ket{11}\}$ is a set of basis vectors for the two-qubit Hilbert
space, the CNOT unitary operator is defined to be 
$ U_{\text{CNOT}}\equiv\ket{00}\bra{00}+\ket{01}\bra{01}+\ket{10}\bra{11}+
 \ket{11}\bra{10}$.
We can now diagonalise CNOT with respect to the basis $\{\ket{00},
\ket{01}, \ket{1+}, \ket{1-}\}$, where
$\ket{\pm}=(|0\rangle\pm|1\rangle)/\sqrt{2}$. The corresponding
eigenvalues are $\{\exp({i0}), \exp({i0}), \exp({i0}),
\exp({i\pi})\}$ respectively. $U_{\text{CNOT}}$ can be written in terms
of its generator $G_{\text{CNOT}}$ as $U_{\text{CNOT}}\equiv e^{i\pi
G_{\text{CNOT}}}=e^{i\pi \ket{1-}\bra{1-}}$, which demonstrates that the
generator of CNOT has only one non-vanishing eigenvalue.
\end{references}
\end{document}